\documentclass[prx,twocolumn, amsmath,amssymb,
reprint,%
author-year,%
author-numerical,%
dvipdfmx,
superscriptaddress
]{revtex4}

\usepackage{graphicx}
\usepackage{bm}
\usepackage{color}

\begin{document}


\title{Small atoms fall into bulk from non-close-packed surfaces?}

\author{Shota Ono}
\email{shota\_o@gifu-u.ac.jp}
\author{Honoka Satomi}
\affiliation{Department of Electrical, Electronic and Computer Engineering, Gifu University, Gifu 501-1193, Japan}
\author{Junji Yuhara}
\affiliation{Graduate School of Engineering, Nagoya University, Nagoya 464-8603, Japan}

\begin{abstract}
Surface rippling has been observed when atoms of $X$ and $A$ are mixed on the $A$ substrate surface. The rippling amplitude has been estimated using hard sphere models. We present a gedanken experiment predicting a penetration of small atoms into bulk through the (100) surface. To understand how the electronic effects alter this picture, we investigate the surface rippling of $X/A(100)$ from first-principles, assuming $X=$ H to Bi except for noble gases and $A=$ Cu, Ag, and Au. We show that the small atoms (such as H, C, N, O and F) attract electrons from the substrate due to the large electronegativity, which prevent them from passing through the void in the (100) surface. The behaviors of small atoms are further explored by studying lateral displacements of the top layer in the $A$ substrate and a formation of the $X$ dimer above, below, and across the top layer. The present work provides an example to understand when atoms are not hard spheres.  
\end{abstract}

\maketitle

\section{Introduction}
In solid state physics, atoms are often treated as hard spheres. For example, a close packing of equally sized spheres explains the crystal structure of simple metals; an assembly of hard spheres shows a crystallization that is known as the Alder transition \cite{alder}; and a hard sphere model provides rationale for the phase transition between B1 (NaCl-type) and B2 (CsCl-type) structures in ionic crystals \cite{AM}. More recently, hard sphere models have been used to understand stable structures in ceramic and refractory materials \cite{smirnov}, superconductors \cite{zhang_sc}, binary and ternary systems \cite{ozaki1,ozaki2}, and two-dimensional ionic crystals \cite{ono_ion}. It has also played an important role in classification of materials \cite{saad}. The advantage of the hard sphere model is its simplicity. On the contrary, it needs to be clarified when the hard sphere model breaks down.  

Overbury and Ku observed the surface rippling of Sn atoms on Cu(111), Ni(111), and Pt(111) surfaces, where they form the $(\sqrt{3}\times\sqrt{3})R30^\circ$ structure and the Sn atoms buckle outward from the first layer \cite{overbury}. The amplitude of the surface rippling decreases with the lattice constant of substrate metals, which has been attributed to the atomic size mismatch between the Sn and substrate atoms. Since the seminal work of them, the surface alloy has served as a playground for investigating the validity of hard sphere models \cite{overbury,wuttig,woodruff,harrison,kimura,xun}. The library of surface alloys has been expanding \cite{Bi_Au111,osiecki,lee,sadhukhan,shah,Ge_Ag111}, and the surface alloy has played an important role in creating or identifying two-dimensional materials \cite{yuhara,pzhang}. 

The description of surface alloys within the hard sphere approximation is based on several factors. First, it depends on the definition of atomic radii such as Clementi's (isolated atom), metallic, and covalent radii \cite{clementi,wells}. The hard sphere model using metallic radii has been known to overestimate the magnitude of rippling amplitude \cite{woodruff}, which has been attributed to a lack of interatomic interaction between the foreign atom and the substrate atom well below the surface \cite{xun}. Second, it depends on the combination of atomic species for the foreign and substrate atoms. Recently, the rippling amplitude of 15 atoms, which are taken from group III, IV, and V in the periodic table, on the Ag surface has been investigated from first-principles \cite{ishii}. This search is, however, restricted to the Ag(111) surface. Third, it depends on the surface orientation of the substrate. It has been suggested that the rippling amplitude is not sensitive to the orientation of the substrate \cite{xun,li}. However, a few combination of atomic species, such as Sn on Pt(111) and Pt(100) \cite{li}, was investigated. In this paper, we point out, through a gedanken experiment, a fundamental problem regarding the applicability of the hard sphere model to the (100) surface, present density-functional theory (DFT) calculations highlighting the effects beyond the hard sphere approximation, and shed light on a peculiarity of small radius elements including H, B, C, and N for surface alloys. 

\section{Gedanken experiment with hard sphere model}
\label{secGed}
\begin{figure}
\center
\includegraphics[scale=0.4]{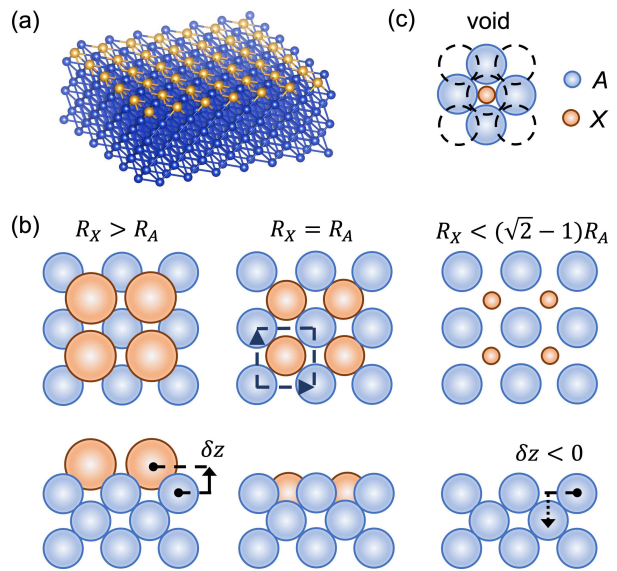}
\caption{(a) Schematic structure of $X/A(100)$ surface alloy with the first six layers, where the $A$ substrate atoms are alloyed with the $X$ foreign atoms to form the c$(2\times 2)$ structure. This is illustrated using \texttt{VESTA} \cite{vesta}. (b) Top and side views of the surface alloy within the hard sphere model. The unit cell is represented by the dashed line. When the atomic radii satisfy the relation $R_X > R_A$, $R_X = R_A$, and $R_X<R_A$, the atom displacement $\delta z$ is positive, zero, and negative, respectively. When $R_X$ is small enough to satisfy the relation of $R_X < (\sqrt{2}-1)R_A$, the $X$ atom can pass through any voids in the square lattice. (c) The void in the second layer. The dashed circle indicates the atomic sphere in the third layer. } \label{fig_1} 
\end{figure}

Let us consider a surface alloy, which consists of $X$ foreign atoms on $A$ substrate in the face-centered cubic (fcc) structure, forming a c$(2\times 2)$ structure on the $A(100)$ surface, i.e., two interpenetrating square lattices of $X$ and $A$, displaced along the diagonal of the square cell by a half length of the diagonal (see Fig.~\ref{fig_1}(a)). With a hard sphere model, we denote the atomic radii of the $X$ and $A$ atoms by $R_X$ and $R_A$, respectively. When $R_A=R_X$, the atoms in the surface layer have no corrugation. When $R_X> R_A$, the $X$ atom is pushed up from the surface, as shown in Fig.~\ref{fig_1}(b). The magnitude of the displacement $\delta z$ from the flat surface is
\begin{eqnarray}
 \delta z = \sqrt{(R_A+R_X)^2 -(2R_A)^2}.
 \label{eq:RX+}
\end{eqnarray} 
When $R_X< R_A$, the $X$ atom is embedded into the first and second layers, yielding a negative value of $\delta z$ expressed by
\begin{eqnarray}
 \delta z = - \sqrt{2}R_A + \sqrt{(R_A+R_X)^2 -2{R}_{A}^{2}}.
 \label{eq:RX-100}
\end{eqnarray} 
It should be noted that Eq.~(\ref{eq:RX-100}) breaks down when the atomic radius is small enough to satisfy the following inequality
\begin{eqnarray}
 R_X < (\sqrt{2}-1) R_A.
\label{eq:break}
\end{eqnarray}
In this case, the $X$ atom passes through the void created in the square lattice of the second layer (see Fig.~\ref{fig_1}(c)), collides with the face-centered atom in the third layer, and eventually falls into the bulk by passing through the void in deep layers. If the $X$ atom is trapped to the third layer (i.e., assuming no displacements parallel to the surface), the displacement is expressed by
\begin{eqnarray}
 \delta z = - a + R_A +R_X,
 \label{eq:third}
\end{eqnarray} 
where $a$ is the lattice constant of $A$ in the fcc structure and thus $z=-a$ is the $z$-component of the third layer of the $A(100)$ surface.

Table \ref{table1} lists the Clementi's atomic radii $R_{\rm C}$ \cite{clementi} for $X=$ H and Li to F and $A=$ Cu, Ag, and Au. These values are derived from solving the Hartree-Fock equation for an isolated atom. We extract the values of $R_{\rm C}$ by using \texttt{pymatgen} \cite{pymatgen}. The condition of Eq.~(\ref{eq:break}) is satisfied when $X=$ H, C, N, O, and F (except for $X=$ C and $A=$ Cu). This clearly shows that the atoms with small atomic radii fall into the bulk through the (100) surface of noble metals. 

The magnitude of the atomic radii depends on the definition. We derive the metallic radii $R_{\rm M}$ from the magnitude of $a$ of $X$ in the fcc structure (i.e., 12-coordination): $R_{\rm M}=\sqrt{2}a/4$. The optimization of $a$ is done by using \texttt{Quantum ESPRESSO} (\texttt{QE}) \cite {qe}, where the computational details are the same as those used in the slab model calculations described below, but a 24$\times$24$\times$24 $k$ grid is used in the self-consistent field (scf) calculations \cite{MK}. When $R_{\rm M}$ is used (see also Table \ref{table1}), the condition of Eq.~(\ref{eq:break}) is never satisfied even for $X=$ H. We should keep in mind the following tendency: When $X$ is changed from B to F, $R_{\rm M}$ deviates around $1.1$ but $R_{\rm C}$ decreases from 0.87 to 0.42. This implies that for $X=$ B to F the coordination has a strong impact on the electronic distribution of valence electrons and enlarges the atomic radius, which may prevent the $X$ atom from passing through the void in the square lattice. 

It is useful to investigate to what extent the electron charge is distributed between different atomic species. We use the Pauling's electronegativity $\chi_{\rm P}$ for atoms of $X$ and $A$ to study which atoms attract electrons strongly, and in Table \ref{table1} list the values of $\chi_{\rm P}$ that are extracted from \texttt{pymatgen} \cite{pymatgen}. The $\chi_{\rm P}$ increases monotonically as one goes from Li to F, so that the $\chi_{\rm P}$ of B to F becomes larger than that of $A=$ Cu, Ag, and Au (except for $X=$ B and $A=$ Au). The effective radius will be larger than $R_{\rm C}$, implying that the atoms of $X$ are trapped near the $A(100)$ surface. This needs to be studied in detail within DFT. 

\begin{table}
\begin{center}
\caption{The Clementi's and metallic atomic radii, $R_{\rm C}$ and $R_{\rm M}$, in units of \AA \ and the Pauling's electronegativity $\chi_{\rm P}$. The figures in a parenthesis is $(\sqrt{2}-1)R$ with $R=R_{\rm C}$ and $R_{\rm M}$ for Cu, Ag, and Au. }
{
\begin{tabular}{lcccccccc}\hline
  \hspace{1mm} & H \hspace{1mm} & Li \hspace{1mm} & Be \hspace{1mm} & B \hspace{1mm} & C \hspace{1mm} & N \hspace{1mm} & O \hspace{1mm} & F  \\ \hline
 $R_{\rm C}$ \hspace{1mm} & 0.53 \hspace{1mm} & 1.67 \hspace{1mm} & 1.12 \hspace{1mm} & 0.87 \hspace{1mm} & 0.67 \hspace{1mm} & 0.56 \hspace{1mm} & 0.48 \hspace{1mm} & 0.42  \\ 
 $R_{\rm M}$ \hspace{1mm} & 0.80 \hspace{1mm} & 1.53 \hspace{1mm} & 1.09 \hspace{1mm} & 1.01 \hspace{1mm} & 1.04 \hspace{1mm} & 1.10 \hspace{1mm} & 1.11 \hspace{1mm} & 1.08 \\ 
 $\chi_{\rm P}$ \hspace{1mm} & 2.20 \hspace{1mm} & 0.98 \hspace{1mm} & 1.57 \hspace{1mm} & 2.04 \hspace{1mm} & 2.55 \hspace{1mm} & 3.04 \hspace{1mm} & 3.44 \hspace{1mm} & 3.98  \\ \hline 
  \hspace{1mm} & Cu \hspace{1mm} &  \hspace{1mm} & Ag \hspace{1mm} &  \hspace{1mm} & Au \hspace{1mm} &  \hspace{1mm} \\ \hline
 $R_{\rm C}$ \hspace{1mm} & 1.45 \hspace{1mm} & (0.60) \hspace{1mm} & 1.65 \hspace{1mm} & (0.68) \hspace{1mm} & 1.74 \hspace{1mm} & (0.72) \\ 
 $R_{\rm M}$ \hspace{1mm} & 1.28 \hspace{1mm} & (0.53) \hspace{1mm} & 1.47 \hspace{1mm} & (0.61) \hspace{1mm} & 1.47 \hspace{1mm} & (0.61) \\ 
 $\chi_{\rm P}$ \hspace{1mm} & 1.90 \hspace{1mm} &  \hspace{1mm} & 1.93 \hspace{1mm} & \hspace{1mm} & 2.54 \hspace{1mm} & \\ 
\hline
\end{tabular}
}
\label{table1}
\end{center}
\end{table}

\begin{figure}
\center
\includegraphics[scale=0.4]{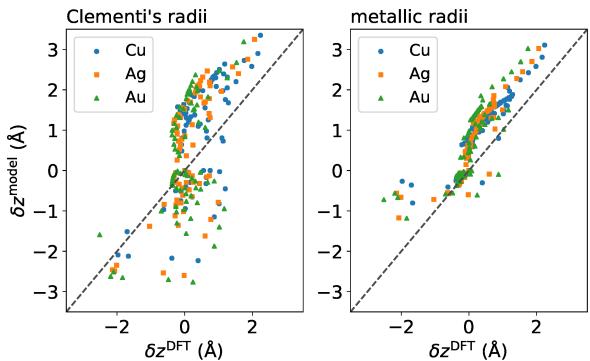}
\caption{Comparison between the atomic displacements for $X/A(100)$ within DFT ($\delta z^{\rm DFT}$) and those within the hard sphere model ($\delta z^{\rm model}$) using Clementi's atomic radii (left) and metallic radii (right). When Eq.~(\ref{eq:break}) is satisfied, Eq.~(\ref{eq:third}) is used. The dashed line indicates $\delta z^{\rm DFT}=\delta z^{\rm model}$.} \label{fig_2} 
\end{figure}

\begin{figure*}
\center
\includegraphics[scale=0.55]{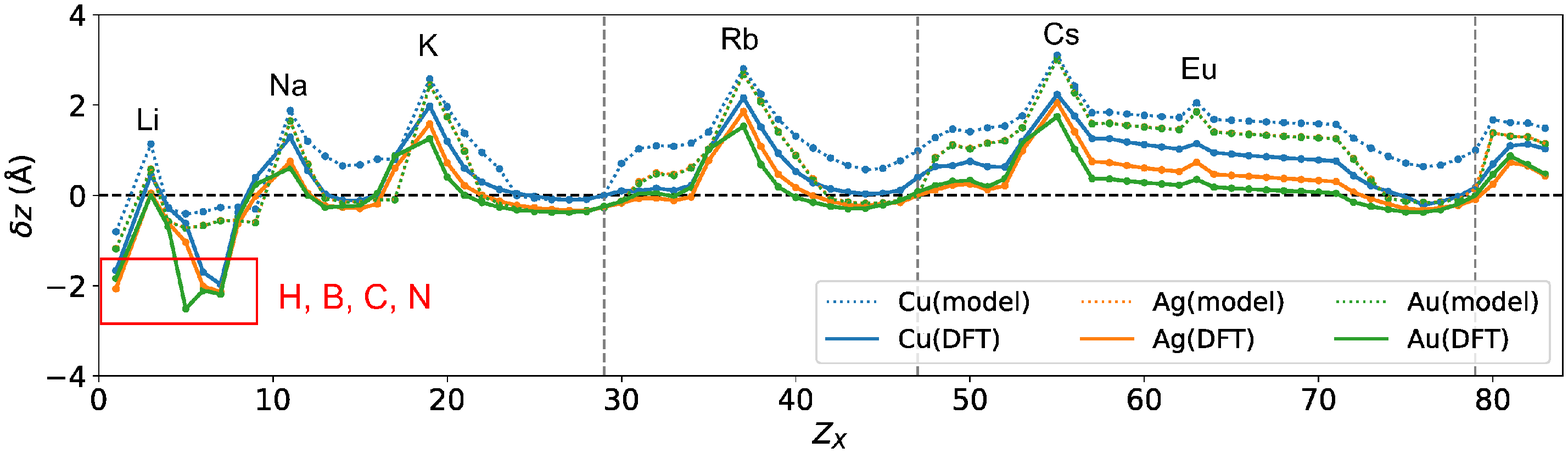}
\caption{The atomic number $Z_X$ dependence of $\delta z$ for the $A(100)$ surface within the DFT and hard sphere model using the metallic radii of $R_{\rm M}$ with $A$ being Cu, Ag, and Au. The vertical lines (dashed) indicate $Z_X=29$ (Cu), $47$ (Ag), and $79$ (Au). Within the hard sphere model, the curves for the Ag and Au almost overlap due to the similar values of $R_{\rm M}$. } \label{fig_3} 
\end{figure*}

\section{DFT calculations}
\label{secDFT}
We use an electronic structure calculation program of \texttt{QE} \cite{qe} to perform DFT calculations for thin films. The metallic substrate $A$ with the (100) surface has 11 atomic layers including 22 atoms. We use the Perdew-Burke-Ernzerhof (PBE) \cite{pbe} functionals of the generalized gradient approximation for the exchange-correlation energy and use the ultrasoft pseudopotentials provided in \texttt{pslibrary.1.0.0} \cite{dalcorso}. The cutoff energies for the wavefunction and the charge density are 55 Ry and 550 Ry, respectively. A 20$\times$20$\times$1 $k$ grid is used in the scf calculations \cite{MK}. The smearing parameter of 0.02 Ry within Methfessel and Paxton approach is used \cite{smearing_MP}. The vacuum layer is taken to be 15 \AA. The interlayer spacing along the [100] direction is fixed to $a/2=1.82, 2.06$, and $2.08$ \AA \ for Cu, Ag, and Au, respectively. The $z$-component of atoms in the first and last three layers (i.e., 1-3 and 9-11th layers) are relaxed in the geometry optimization. For the first and eleventh layers, we next replace the $A$ atom with the $X$ atom to form the c$(2\times 2)$ structure, still having an inversion symmetry in the model. We again optimize the $z$-component of the $X$ and $A$ atoms within the first and last three layers, i.e., an initial guess of $\delta z = 0$ \AA. In the present work, we study $A=$ Cu, Ag, and Au, allowing us to understand a trend in the group 11 elements. The atom of $X$ ranges from H to Bi, where noble gases are excluded. The magnetic effect on $\delta z$ will be important especially for $X=$ Mn \cite{wuttig,harrison}, but is not investigated in the present work. The parameters of $R_{\rm C}$, $R_{\rm M}$, $a$, and $\chi_{\rm P}$ for $X$, and the calculated values of $\delta z$ for $X/A(100)$ are provided in Supplemental Material \cite{SM}. 

\subsection{Breakdown of hard sphere model}
Figure \ref{fig_2} shows a relationship between $\delta z$s calculated within DFT and those within hard sphere models using $R_{\rm C}$ (left) and $R_{\rm M}$ (right). When $R_{\rm C}$ is used, the agreement is not good because the calculated data are distributed around the vertical line of $\delta z^{\rm DFT}=0$. On the other hand, when $R_{\rm M}$ is used, $\delta z^{\rm model}$ is well correlated with $\delta z^{\rm DFT}$. This is consistent with experiments \cite{overbury}, where a positive value of $\delta z$ is caused by the lattice constant mismatch between $X$ and $A$ (i.e., $a=4R_{\rm M}/\sqrt{2}$). This result implies that the atomic environment near the surface is similar to  that in the fcc structure. We also find that $\delta z^{\rm model}$ is larger than $\delta z^{\rm DFT}$. Such an overestimation has also been reported in Refs.~\cite{woodruff,xun}, while it has been remedied by including long-range forces between hard spheres \cite{xun}. Below the line of $\delta z^{\rm DFT}= \delta z^{\rm model}$, we observe some anomalies corresponding to F/$A(100)$ and Cl/$A(100)$ except for Cl/Cu(100). This is due to the large electronegativity of F and Cl, which will be discussed below. 
 
Figure \ref{fig_3} shows $\delta z$ as a function of the atomic number $Z_X$ of the $X$ atom on noble metal substrates. The trend for $\delta z$ as a function of $Z_X$ is similar between the model and DFT calculations: (i) The magnitude of $\delta z$ is maximum for the alkali atoms of $X$, which is due to the small $Z_X$ in alkali metals (group I), yielding an electron delocalization around the nucleus and an enhancement of atomic radius; (ii) $\delta z$ for $A=$ Cu is larger than that for $A=$ Ag and Au, which may be due to the large atomic radii of Ag and Au compared with Cu atom; and (iii) a peak of $\delta z$ is observed at $Z_X=63$, europium (Eu). The Eu crystal has the body-centered cubic structure as the ground state, which is similar to alkali metals, while the other lanthanides have closed-packed structures. In this way, the hard sphere model can be used to interpret DFT results. 

Let us move on to the cases of $X=$ H and Li to F. When $X=$ H, C and N, a negatively large value of $\delta z$ is observed. This indicates that the $X$ atoms are trapped to the second layer from the surface (i.e., $\delta z\simeq - a/2$). We also observe anomalous decrease in $\delta z$ for B/Au(100). This is not described by the hard sphere model using $R_{\rm M}$. When $X=$ O and F, a small negative or positive value of $\delta z$ is obtained. This is not described by the hard sphere model using $R_{\rm C}$. In this way, no consistent picture can be obtained within hard sphere models, and thus the electronic property beyond the hard sphere approximation is investigated below by studying the case of $A=$ Cu.

\begin{figure*}
\center
\includegraphics[scale=0.5]{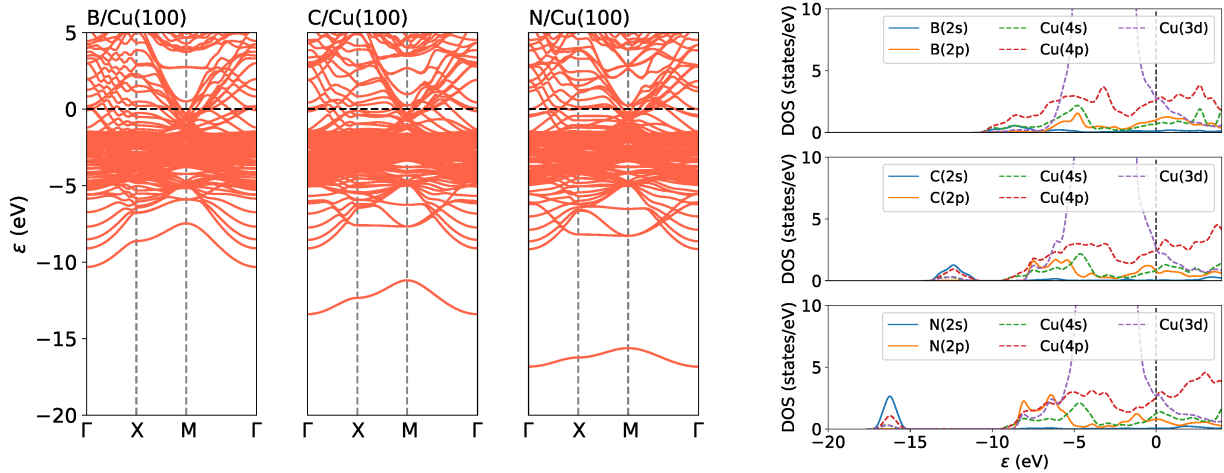}
\caption{Electronic band structure (left) and projected density-of-states (right) for B/Cu(100), C/Cu(100), and N/Cu(100). The electron energy $\varepsilon$ is measured from the Fermi level.} \label{fig_4} 
\end{figure*}

Note that we have calculated the potential energy as a function of $\delta z$ by assuming $X=$ H and Li to F on Cu(100) surface. We observed parabola for each $X$, confirming no dependence of the initial condition for $\delta z$. This is provided in Supplemental Material \cite{SM}. We have also studied the van der Waals effect on $\delta z$ and performed DFT calculations including a Grimme's van der Waals correction (DFT-D3) \cite{grimme}, assuming $X=$ H and Li to F and $A=$ Cu, Ag, and Au. Such calculations also predict a trap to the second layer of the surface, while the value of $\delta z$ increases by less than 0.1 \AA. In addition, we have studied the energetic stability of $X/A(100)$ by calculating the formation energy $\Delta E$ and demonstrated that the $X/A(100)$ is stable for $X=$ B, C, and N, confirming a correlation between $\Delta E$ and $\delta z$ (except for $X=$ H). This is also provided in Supplemental Material \cite{SM}. 

We emphasize that no atomic trap to the second layer is observed for the $A(111)$ surface in the $(\sqrt{3}\times\sqrt{3})R30^\circ$ structure. Within the hard sphere model, the rippling is expressed by $\delta z = - \sqrt{8/3}R_A + \sqrt{(R_A+R_X)^2 -4{R}_{A}^{2}/3}$ when $R_X< R_A$ \cite{ishii}. This expression breaks down when $R_X<(2/\sqrt{3}-1)R_A\simeq 0.15 R_A$, whereas the factor of $0.15$ is very small. Although the B, C, and N atoms take negative value of $\delta z$ within the DFT, they are located in between the first and second layers rather than trapped to the latter. Note that the H atom is close to the second layer because $\delta z=-1.6$ and $-1.5$ \AA \ for $A=$ Ag and Au, respectively. The $Z_X$-dependence of $\delta z$ for $A(111)$ surface is compared with that for $A(100)$ surface in Supplemental Material \cite{SM}.

\begin{figure*}
\center
\includegraphics[scale=0.55]{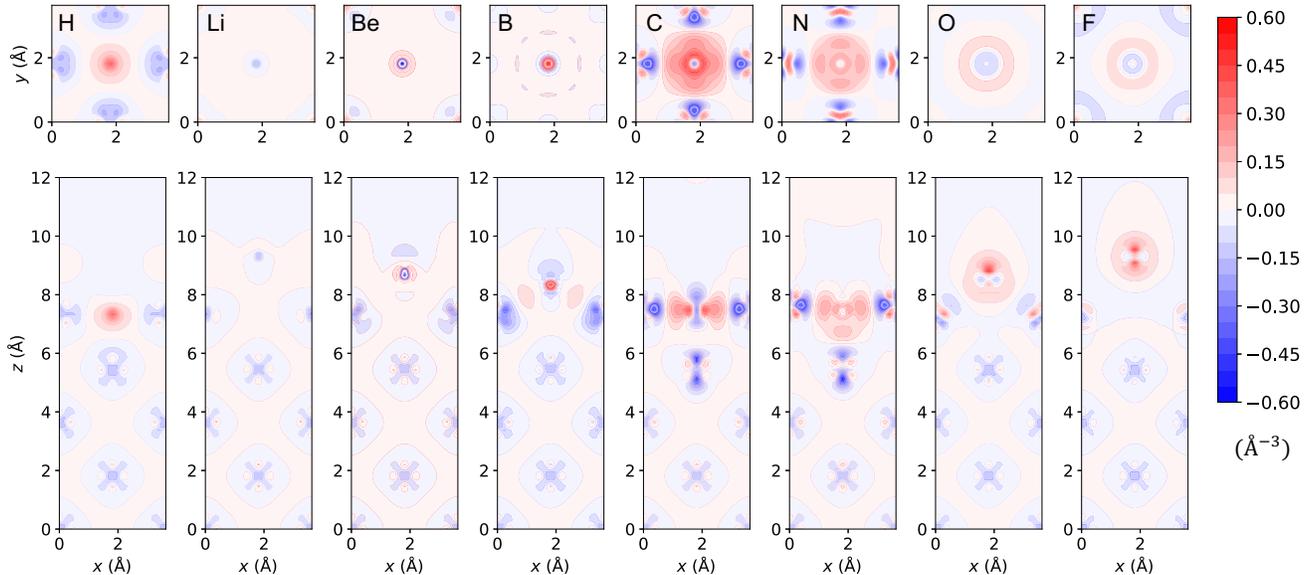}
\caption{Total electron density (\AA$^{-3}$) subtracted by the superposition of electron density of atoms. The $x$-$y$ (top) and $x$-$z$ (bottom) planes including the $X$ atom are shown for $X$/Cu(100) with $X=$ H and Li to F. The positive (red) and negative (blue) values indicate an increase and decrease in the electron density compared to that of the isolated atom, respectively. } \label{fig_5} 
\end{figure*}

\subsection{Electronic effect}
Figure \ref{fig_4} shows the electron band structure and projected density-of-states (PDOS) for $X=$ B, C, and N on Cu(100). As can be seen in a Cu bulk, we observe parabolic bands with the minimum electron energy at the $\Gamma$ point and non-dispersive bands from $\varepsilon=-5$ to $-1.5$ eV, which originate from the Cu $4s$ and $3d$ electrons, respectively. In addition, a cosine-type band is located well below the Fermi level. As the $Z_X$ is increased, the location of the band becomes deep. From the PDOS, we identify that such a cosine band originates from the $X$ $2s$ and Cu $4p$ electrons. We also find that the $X$ $2p$ electron contributes to the PDOS for $\varepsilon \ge -9$ eV, implying that $X$ $2p$ orbital hybridizes with Cu orbitals. 

Figure \ref{fig_5} shows a difference between the total electron density and the superposition of atomic electron density, assuming the Cu(100) surface. The electron transfer occurs from Cu to $X$ atoms except for $X=$ Li and Be, which is consistent with the electronegativity analysis using Table \ref{table1}. For $X=$ C and N, anisotropic distribution in the electron density is observed around the second layer of the surface: The electron density around the $X$ atom increases within the $x$-$y$ plane, which gives rise to electron redistribution of Cu atoms in the second ($z=1.82\times 4$ \AA) as well as the third ($z=1.82\times 3$ \AA) layers. The electron redistribution for $X=$ H is spherically symmetric, compared to that for $X=$ C and N, which is intrinsic to the $1s$ orbital. For $X=$ O and F, the effective atomic radius is large due to the large $\chi_{\rm P}$, explaining no penetration into a bulk. 

\begin{figure}
\center
\includegraphics[scale=0.5]{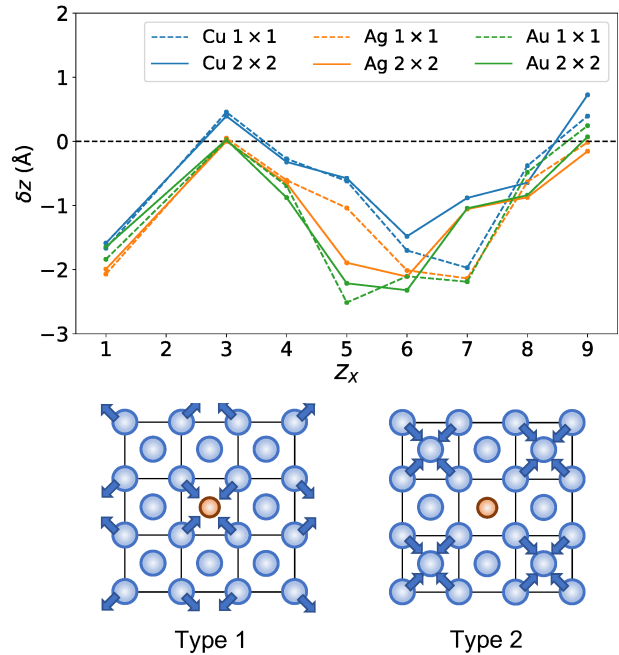}
\caption{(Top) The $\delta z$ for $A(100)$ surface with $X=$ Li to F and $A=$ Cu, Ag, and Au. $1\times 1$ (dashed) and $2\times 2$ (solid) supercells are assumed, where the former results are the same as those obtained for c$(2\times 2)$ calculations. (Bottom) The atomic displacements obtained by the $2\times 2$ supercell calculations. The $A$ atoms (blue) approach to and go away from the $X$ atom (orange) for type 1 and type 2, respectively.  } \label{fig_6} 
\end{figure}

\subsection{Atomic density dependence}
\label{secDis}
We study to what extent the lateral force within the surface (i.e., the periodic boundary condition along the $x$- and $y$-directions) is important for yielding negatively large values of $\delta z$. We have performed $2\times 2$ supercell calculations, where the surface of the slab consists of seven $A$ atoms and one $X$ atom in the unit cell and 88 atoms are present in the slab. The $x$- and $y$- as well as $z$-components of atoms are relaxed in the first and last three layers. A 5$\times$5$\times$1 $k$ grid is used in the scf calculations \cite{MK}. We still observe a trap around the second layer when $X=$ H, B and C, except for B/Cu(100), as shown in Fig.~\ref{fig_6}(top). It is interesting that the atomic distribution is classed to two types illustrated in Fig.~\ref{fig_6}(bottom): the $A$ atoms are displaced to approach the $X$ atom (type 1) and to go away from the $X$ atom (type 2). The type 1 includes Be/Cu, B/Cu, F/Cu, Be/Ag, and F/Ag, and type 2 otherwise. Only type 2 systems can show negatively large $\delta z$. However, for $X=$ N taking the type 2, the value of $\delta z$ increases by about 1 \AA, so that the N atoms are located in between the first ($\delta z=0$ \AA) and second ($\delta z\simeq 2$ \AA) layers. 

We next explore surface properties specific to the H, B, C, and N atoms. The negatively large $\delta z$ indicates the presence of a deep hole on the (100) surface in the c$(2\times 2)$ structure, which may accommodate more atoms. We have added one $X$ atom to form a $X$ dimer whose symmetry axis is parallel to the $z$ axis and optimized the $z$-components of $X$ and $A$ atoms. We have obtained three patterns for alignment of two $X$ atoms: a $X$ dimer formation above ($\delta z_1, \delta z_2>0$), below ($\delta z_1, \delta z_2<0$), and across ($\delta z_1<0, \delta z_2>0$) the first layer, as listed in Table \ref{table2}. The Ag(100) and Au(100) surfaces push the N dimer away from the surface and an isolated N dimer is formed, indicating that a N dimer is immiscible with Ag and Au surfaces. On the other hand, the Cu(100), Ag(100), and Au(100) surfaces accommodate the N dimer, the H dimer, and the H and B dimers, respectively. The interatomic distance $d$ of the N and B dimers below the first layer is 1.25 and 1.68 \AA, respectively, which is larger than that of the isolated N and B dimers ($d=$ 1.11 and 1.64 \AA \ within DFT-PBE). The $d$ of the H dimer is much larger than that of the isolated $H$ dimer ($d=0.75$ \AA). This implies that two H atoms are trapped to the first and second $A$ layers rather than that a H dimer is formed below the first layer, while for $A=$ Au the H atoms tend to be trapped to the second and third layers. For the other cases, the $X$ dimer is formed across the first layer of $A(100)$ surface. These results suggest that tuning the interatomic distance of $X$ dimer can be possible by changing the substrate. 


\begin{table}
\begin{center}
\caption{The $\delta z$ for two $X$ atoms relative to the first layer of the $A(100)$ surface. $d=\delta z_2 - \delta z_1$ is the interatomic distance of $X_2$. The $d$ for an isolated $X$ dimer is also listed.}
{
\begin{tabular}{lrrc}\hline
 $X_2/A$ \hspace{5mm} & $\delta z_1$ \hspace{5mm} & $\delta z_2$ \hspace{5mm} & $d$ \hspace{5mm}\\ \hline
H$_2$/Cu \hspace{5mm} & $-1.99$ \hspace{5mm} & 0.08 \hspace{5mm} & 2.07 \hspace{5mm}\\
H$_2$/Ag \hspace{5mm} & $-2.39$ \hspace{5mm} & $-0.07$ \hspace{5mm} & 2.32 \hspace{5mm}\\
H$_2$/Au \hspace{5mm} & $-3.18$ \hspace{5mm} & $-1.52$ \hspace{5mm} & 1.66 \hspace{5mm}\\
H$_2$       \hspace{5mm} & - \hspace{5mm} & - \hspace{5mm} & 0.75 \hspace{5mm}\\
B$_2$/Cu \hspace{5mm} & $-1.22$ \hspace{5mm} & 0.37 \hspace{5mm} & 1.59 \hspace{5mm}\\
B$_2$/Ag \hspace{5mm} & $-1.44$ \hspace{5mm} & 0.15 \hspace{5mm} & 1.58 \hspace{5mm}\\
B$_2$/Au \hspace{5mm} & $-3.11$ \hspace{5mm} & $-1.42$ \hspace{5mm} & 1.68 \hspace{5mm}\\
B$_2$       \hspace{5mm} & - \hspace{5mm} & - \hspace{5mm} & 1.64 \hspace{5mm}\\
C$_2$/Cu \hspace{5mm} & $-1.02$ \hspace{5mm} & 0.29 \hspace{5mm} & 1.31 \hspace{5mm}\\
C$_2$/Ag \hspace{5mm} & $-1.07$ \hspace{5mm} & 0.22 \hspace{5mm} & 1.29 \hspace{5mm}\\
C$_2$/Au \hspace{5mm} & $-1.13$ \hspace{5mm} & 0.19 \hspace{5mm} & 1.32 \hspace{5mm}\\
C$_2$       \hspace{5mm} & - \hspace{5mm} & - \hspace{5mm} & 1.28 \hspace{5mm}\\
N$_2$/Cu \hspace{5mm} & $-1.75$ \hspace{5mm} & $-0.50$ \hspace{5mm} & 1.25 \hspace{5mm}\\
N$_2$/Ag \hspace{5mm} & 3.59 \hspace{5mm} & 4.70 \hspace{5mm} & 1.11 \hspace{5mm}\\
N$_2$/Au \hspace{5mm} & 3.57 \hspace{5mm} & 4.68 \hspace{5mm} & 1.11 \hspace{5mm}\\
N$_2$       \hspace{5mm} & - \hspace{5mm} & - \hspace{5mm} & 1.11 \hspace{5mm}\\
\hline
\end{tabular}
}
\label{table2}
\end{center}
\end{table}


\section{Summary}
\label{secSum}
We have presented a thought experiment that the small atoms, $X=$ H and C to F, fall into a bulk through the (100) surface, which is based on a hard sphere model using the Clementi's atomic radii. To check whether this is true, we have performed DFT calculations on the surface rippling for $X/A(100)$ in the c$(2\times 2)$ structure with $A=$ Cu, Ag, and Au. The small atoms stay near the surface rather than penetrate into a bulk: the H, C, and N atoms are trapped to the second layer of the (100) surface, while the O and F atoms stay around the first layer. The effects beyond the hard sphere approximation have been revealed by studying electronic properties of $X/A(100)$ and performing $2\times 2$ supercell calculations. These calculations have suggested an importance of the electronegativity difference between atomic species and also the atomic displacement along the parallel to the surface. We have also studied a formation of $X$ dimer around the $A(100)$ surface and shown that the N atoms tend to form a N dimer, while the H atoms trap to different layers. 

We have exhaustively investigated the surface rippling of $X=$ H to Bi except for noble gases, providing a benchmark for the surface rippling in the c$(2\times 2)$ structure. We demonstrate that (i) a hard sphere model parametrized by metallic radii can capture an overall trend of the rippling amplitude, suggesting that a consideration of atomic environment into atomic sphere is important, and (ii) such a model fails to predict the structure of non-close-packed surfaces including a small atom. The electronic effect overcomes the geometric effect on the structural property. On the contrary, we can expect that surfaces including small radius elements such as H, B, C, and N would have interesting structures beyond the hard sphere description, which potentially show useful physical and chemical properties. In particular, the behavior of H atoms will be of importance in high-temperature superconductivity \cite{zhang_sc} and hydrogen storage \cite{ren}. 

\begin{acknowledgments}
This work was supported by JSPS KAKENHI (Grant No. 21K04628). The computation was carried out using the facilities of the Supercomputer Center, the Institute for Solid State Physics, the University of Tokyo, and also using the supercomputer ``Flow'' at Information Technology center, Nagoya University. 
\end{acknowledgments}



\end{document}